\documentclass[useAMs,usenatbib,a4paper]{mn2e}
\usepackage{savesym}
\usepackage{graphicx}
\expandafter\let\csname equation*\endcsname\relax
  \expandafter\let\csname endequation*\endcsname\relax 
\usepackage{subfig}
\usepackage{amsmath}
\usepackage{amssymb}
\usepackage{verbatim}
\usepackage{array}
\usepackage{times}
\usepackage[total={17.8cm,24.0cm},centering]{geometry} 

\newcommand{\be}{\begin{equation}}
\newcommand{\beq}{\begin{equation}}
\newcommand{\ee}{\end{equation}}
\newcommand{\eeq}{\end{equation}}
\newcommand{\eea}{\end{eqnarray}}
\newcommand{\bea}{\begin{eqnarray}}

\newcommand{\m}{\mathrm}
\newcommand{\Pm}{{\rm Pm}}

\title[Simulations of the Pm disc instability]{Demonstration of a magnetic Prandtl number disc instability from first principles}
\author[William J. Potter and Steven A. Balbus]{William J. Potter\thanks{E-mail:
will.potter@astro.ox.ac.uk (WJP)} and Steven A. Balbus
\\
Oxford Astrophysics. Denys Wilkinson Building, Keble Road, Oxford, OX1 3RH, United Kingdom}
\begin{document}

\date{}

\pagerange{\pageref{firstpage}--\pageref{lastpage}} \pubyear{2017}

\maketitle

\label{firstpage}

\begin{abstract}
Understanding what determines the strength of MHD turbulence in accretion discs is a question of fundamental theoretical and observational importance. In this work we investigate whether the dependence of the turbulent accretion disc stress ($\alpha$) on the magnetic Prandtl number ($\Pm$) is sufficiently sensitive to induce thermal-viscous instability using 3D MHD simulations. We first investigate whether the $\alpha$-$\Pm$ dependence, found by many previous authors, has a physical or numerical origin by conducting a suite of local shearing-box simulations. We find that a definite $\alpha$-$\Pm$ dependence persists when simultaneously increasing numerical resolution and decreasing the absolute values of both the viscous and resistive dissipation coefficients. This points to a physical origin of the $\alpha$-$\Pm$ dependence. Using a further set of simulations which include realistic turbulent heating and radiative cooling, and by giving $\Pm$ a realistic physical dependence on the plasma temperature and density, we demonstrate that the $\alpha$-$\Pm$ dependence is sufficiently strong to lead to a local instability. We confirm that the instability manifests itself as an unstable limit cycle by mapping the local thermal-equilibrium curve of the disc. This is the first self-consistent MHD simulation demonstrating the $\Pm$ instability from first principles. This result is important because a physical $\Pm$ instability would lead to the global propagation of heating and cooling fronts and a transition between disc states on timescales compatible with the observed hard/soft state transitions in black hole binaries.
\end{abstract}


\section{Introduction}

Accretion discs surrounding compact objects exhibit complex and dramatic cyclic changes in X-ray luminosity on a variety of timescales (e.g. \citealt{2004MNRAS.355.1105F}, \citealt{2006ARA&A..44...49R} and \citealt{2007A&ARv..15....1D}). Black hole binaries (hereafter ``BHB'') are observed to undergo flaring and quiescent cycles on month-long timescales in which the X-ray luminosity of the disc changes by orders of magnitude. This long-timescale cyclic flaring is thought to be caused by a thermal-viscous instability due to the sharp increase in disc opacity when hydrogen becomes ionised. This is known as the disc instability mechanism (DIM) (\citealt{1983MNRAS.205..359F}, \citealt{1998MNRAS.298.1048H} and \citealt{2001NewAR..45..449L}). In the last decade, cyclic changes in the disc luminosity and X-ray spectrum have been observed in which the disc cycles between a radiatively efficient high soft state, resembling a classic composite blackbody and a radiatively inefficient low hard state, with a power law non-thermal hard X-ray spectrum. These changes in disc state are of particular interest because they are intimately linked to the production of a relativistic jet in the system; in the hard state a radio jet is observed, whilst in the soft state no jet is observed. However, the physical mechanism responsible for these state changes is not yet understood. In this paper we present simulations of a promising new thermal-viscous instability which could explain the hard/soft changes in disc state.   

In a previous paper \citep{2014MNRAS.441..681P} (hereafter PB), we put forth arguments for the existence of a new type of thermal-viscous disc instability triggered when the disc stress (usually parameterised by $\alpha$, \citealt{1973A&A....24..337S}) depends sensitively on the physical properties of the disc plasma, especially temperature (see also the instability calculations in \citealt{2011ApJ...727..106T}).  Simulations both of accretion disc turbulence  (\citealt{2007A&A...476.1123F},  \citealt {2007MNRAS.378.1471L},  \citealt{2009ApJ...707..833S}, \citealt{2015A&A...579A.117M}), and of driven turbulent dynamos (\citealt{2004ApJ...612..276S}) had shown that when it is near unity, the magnetic Prandtl number (the ratio of the microscopic viscosity to resistivity, $\Pm=\nu/\eta$), can have a significant effect on the strength and the maintenance of turbulent fluctuations.  A possible explanation for the dependence of the disc $\alpha$ parameter on $\Pm$ is that the rate of small scale magnetic reconnection in turbulent flow is sensitive to the strength of the viscous stress in a reconnecting layer when Pm is near unity \citep{2008ApJ...674..408B}.   

Magnetic and kinetic energy are extracted from the differential rotation of the disc by the magnetorotational instability (MRI), predominantly on spatial scales set by the disc scale height \cite{1998RvMP...70....1B}.   Once the MRI is fully non-linear, the gas hosts a turbulent cascade, transferring energy downwards to ever smaller length scales, until finally it is dissipated as heat.   In the  $\Pm>1$ regime, the viscosity is larger than the resistivity on a given scale, and so the viscous dissipation is correspondingly larger than resistive dissipation; the opposite is true when $\Pm<1$.   Both the growth rate of the MRI and the magnitude of the turbulent disc stress (most of which is magnetic) are related to the RMS magnetic field strength of the disc.  The saturated magnetic field strength in turn depends upon the rate of magnetic reconnection in the plasma.   When $\Pm>1$,  the viscous dissipation scale exceeds the resistive scale and so the viscosity damps out velocity fluctuations on the resistive scale.   This leads to a lower rate of reconnection via dissipation, a build-up of the magnetic energy, and an increased value of $\alpha$.   This is because the field dissipation process will generally require large velocity gradients over the resistive lengthscale in order to bring misaligned magnetic field lines together.  If these large gradients produce correspondingly large viscous stresses, the dissipation will be inhibited.   If on the other hand $\Pm<1$, the resistive length scale is larger than the viscous length scale, and velocity gradients on this resistive scale will not produce important dynamical stress.  Magnetic dissipation unencumbered by viscosity will proceed apace, the cascade decreasing the saturated magnetic field strength, and with it the $\alpha$ stress.

The disc radius at which the transition between $\Pm>1$ and $\Pm<1$ occurs is $r\sim100r_{\m{s}}$ in a typical BHB accretion disc (PB).  The dependence of $\alpha$ on $\Pm$ is typically a power law, $\alpha \propto \m{Pm}^{n}$.   PB showed that within the classic $\alpha$ formalism, when $n>0.5$,  the disc is unstable.  Many (though not all) MHD simulations of the dependence of $\alpha$ on $\Pm$ had found $n\sim0.5-1$, suggesting that the instability might plausibly be present in astrophysical discs (\citealt{2007A&A...476.1123F} \citealt{2007MNRAS.378.1471L} \citealt{2009ApJ...707..833S} etc.).  Using an idealised 1D dynamic thin disc approximation, PB demonstrated that the instability does indeed lead to the formation of a local unstable limit cycle.    Dynamical heating and cooling fronts move throughout the disc, as in classical dwarf nova modelling (e.g. \citealt{1998MNRAS.298.1048H}).   

These preliminary results suggest two further avenues of exploration, which we follow here.   The first is to investigate with some care the dependence of the disc stress upon the magnetic Prandtl number in a variety of initial conditions using 3D MHD isothermal shearing-box simulations.   It is obviously critical to establish that the $\Pm$ dependence is physical, and not a numerical artefact.  The second challenge is to conduct simulations of the disc thermal equilibrium curve using more realistic cooling, including a temperature and density dependent Pm.   Does the suspected instability actually occur in a first-principle MHD simulation?   

\section{Numerical setup}

We carry out this investigation with the PLUTO MHD code \citep{2007ApJS..170..228M}, which is widely-used and publicly available. The dissipative MHD equations are solved in the local shearing-box approximation using a Godunov-type finite volume scheme with explicit dissipation terms 
(\citealt{2008A&A...487....1B}, \citealt{1995ApJ...440..742H}, \citealt{2012A&A...545A.152M}, \citealt{2014ApJ...787L..13B}).  In conservation form, the equations are:  
\be
\frac{\partial \rho}{\partial t}+\nabla\cdot(\rho {\bf v})=0,
\ee
\be
\frac{\partial (\rho {\bf v})}{\partial t}+\nabla\cdot(\rho{\bf vv}-{\bf BB})+\nabla P_{\m{tot}}=\rho{\bf g}_{s} - 2\Omega_{0}{\bf \hat{z}}\times\rho{\bf v}+\nabla\cdot\Pi,
\ee
\bea
\frac{\partial E}{\partial t}+\nabla\cdot[(E+P_{\m{tot}}){\bf v}-({\bf v\cdot B}){\bf B}]&=&\nonumber \\ &&\hspace{-4.5cm}...\rho{\bf v}\cdot{\bf g}_{s}-\nabla\cdot[(\eta\cdot{\bf J})\times{\bf B}]+\nabla\cdot({\bf v}\cdot \Pi)-\Lambda,
\eea
\be
\frac{\partial {\bf B}}{\partial t}-\nabla\times({\bf v}\times{\bf B})=-\nabla\times(\eta\cdot{\bf J}),
\ee
\be
{\bf J}=\nabla\times{\bf B}, \qquad P_{\m{tot}}=P+\frac{B^{2}}{2},
\ee
\be
E=\rho e+\frac{\rho v^{2}}{2}+\frac{B^{2}}{2}, \qquad  {\bf g}_{s}=\Omega_{0}^{2}(2qx{\bf \hat{x}}-z{\bf\hat{z}}),
\ee
\be
\qquad \Pi_{ij}=\nu\left[\frac{\partial v_{i}}{\partial x_{j}}+\frac{\partial v_{j}}{\partial x_{i}}-\frac{2}{3}\nabla\cdot{\bf v}\delta_{ij}\right].
\ee
where $\rho$ is the mass density, ${\bf v}$ the fluid velocity, ${\bf B}$ the magnetic field, $P_{\m{tot}}$ the total pressure (thermal plus magnetic), $P$ the thermal pressure, ${\bf g}_{s}$ the effective gravity in the shearing-box approximation, $\Omega_{0}$ is the Keplerian angular velocity at the centre of the shearing-box, $\Pi$ the viscous stress tensor, $E$ the energy density, $e$ the internal energy per unit mass, $\eta$ the resistivity, $\nu$ the shear viscosity (we assume $\eta$ and $\nu$ are diagonal and isotropic tensors, i.e. scalars), ${\bf J}$ the electric current density and $\Lambda$ the energy loss per unit volume due to radiative cooling. 

\subsection{Cooling}

We implement an effective cooling function in the simulation assuming that the opacity is dominated by electron scattering.  This is likely to be the case in the inner regions of the disc, where we expect the $\Pm=1$ transition to occur (PB). Following \citealt{1983MNRAS.205..359F} and \citealt{2012MNRAS.426.1107L}, we use a bulk cooling function to stand in for the diffusive cooling of the form
\be
\Lambda=\frac{\sigma T_{e}^{4}}{H},
\ee
where $H$ is the disc scale height $H=c_{s}/\Omega$, $c_{s}$ is the thermal sound speed i.e. $P=\rho c_{s}^{2}$, $\Omega$ is the Keplerian angular velocity and $T_{e}$ is the effective surface temperature given by
\be
T_{e}^{4}=\frac{4T^{4}}{3\Sigma\kappa}.
\ee
$T$ is the temperature of the simulated central disc plasma, $\kappa=0.4\m{cm}^{2}\m{g}^{-1}$ is the electron scattering opacity and $\Sigma=\rho H$ is the disc surface density. The cooling is then calculated using the temperature and density in each simulation cell. 

\subsection{Instability criterion and $\Pm$}

In PB we derived a thermal-viscous instability criterion for a standard, radiatively efficient thin disc in which the $\alpha$ parameter is a variable depending on $\rho$ and $T$.   This is
\be
\frac{\partial \ln \alpha}{\partial \ln \Sigma}+\frac{1}{4}\frac{\partial \ln \tau}{\partial \ln \Sigma} + 1<0,
\label{geninst}
\ee
where $\tau=\Sigma\kappa$ is the optical depth of the disc. Assuming a power-law dependence for $\alpha$ upon the magnetic Prandtl number, $\alpha \propto \Pm^{n}$, a parametisation suggested by simulations (\citealt{2007MNRAS.378.1471L}, \citealt{2009ApJ...707..833S}), the disc is unstable if $n>0.5$  (PB, equations 46 and 47).  Here $\alpha$ is determined by
the usual formula
\be
\alpha P_{\m{tot}}=\langle \rho(\delta v_{r} \delta v_{\phi}- v_{A\,r}v_{A\,\phi})\rangle,
\ee
where $\delta v_{x}$ is residual velocity after subtraction of the local Keplerian circular velocity (i.e. $\delta v_{\phi}=v_{\phi}-\Omega R$, $\delta v_{r}=v_{r}$), $v_{A\,x}$ is the Alfv\'{e}n velocity in the $x$-direction, $v^{2}_{A\,x}=B_{x}^{2}$, and the angle brackets denote a spatial average. For a typical BHB accretion disc environment, the radiative viscosity dominates the Coulomb viscosity, and one finds:
\be
\eta=\frac{5.6\times10^{11}\ln \Lambda_{eH}}{T^{3/2}} \m{cm}^{2}\m{s}^{-1}, \,\,\, \nu_{\m{Rad}}=\frac{6.7\times10^{-26}T^{4}}{\kappa \rho^{2}} \m{cm}^{2}\m{s}^{-1},
\ee
\be
\Pm\simeq\frac{\nu_{\m{Rad}}}{\eta}=1.9\times10^{-38}\frac{T^{11/2}}{\kappa\rho^{2}},
\label{Pm}
\ee
where $\kappa$ is the opacity of the plasma and $\ln\Lambda_{eH}$ is the electron-proton Coulomb collision factor, which we take to have a value of $\sqrt{40}\simeq6.3$ (see PB section 2.1.3 for more detail).

\section{How does $\alpha$ depend on dissipation and resolution?}

In this section we wish to answer two questions, (i) to what extent do the viscous and resistive dissipation coefficients determine the strength of the saturated disc turbulence quantified by $\alpha$; and (ii) to what extent are these effects physical or numerical in origin? Recent simulations from several groups show that $\alpha$ depends on a whole array of physical and numerical parameters, such as the height of the simulating box, stratification, the presence of net magnetic fields, numerical resolution and convection, etc.  (e.g., \citealt{2012MNRAS.422.2685S}, \citealt{1995ApJ...440..742H}, \citealt{2007A&A...476.1123F}, \citealt{2014ApJ...787....1H} and \citealt{2017arXiv170200777R}).  Here we are interested in isolating the effect of dissipation coefficients.   To avoid conflating this with other complications, we initially choose the simplest isothermal, unstratified local shearing-box.  We wish to study the effect of dissipation coefficients, numerical resolution and the initial magnetic field configuration.  In these isothermal simulations we maintain fixed dissipation coefficients throughout an individual run.  (We shall later allow the viscosity and resistivity to become time-dependent functions of temperature and density.)

In Figures \ref{fig1}-\ref{fig3} and Table \ref{Table1} we summarise the results of the isothermal simulations for a variety of values of resistivity, viscosity, net magnetic field and resolution. The resistivity and viscosity can be expressed in terms of the dimensionless Reynolds number, $\rm{Re}$, and magnetic Reynolds number Rm, given by.
\be
\m{Rm}=\frac{c_{s}H}{\eta}, \qquad \m{Re}=\frac{c_{s}H}{\nu}, \qquad \Pm=\frac{\m{Rm}}{\m{Re}}.
\ee 
\begin{table*}
\centering
\begin{tabular}{| c | c | c | c | c | c | c | c | c | c |}
\hline
Simulation set & $\beta_{\phi}$ & $\beta_{z}$ &$ \Pm$ & Rm & Resolution (x,y,z) & E.O.S & $\rho/\rho_{0}$ & $\kappa$ & Pm$_{\m{initial}}$ \\ \hline
IsoBphi1 & $100$ & n.a. & 1/2, 1, 2, 4 & variable & $64\times100\times64$ & isothermal & n.a. & n.a. & n.a.\\ \hline 
IsoBphi2 & $100$ & n.a. & 1/16, 1/4, 1, 4, 16 & 12800 & $64\times100\times64$ & isothermal & n.a. & n.a. & n.a.\\ \hline 
IsoBphi3 & 1000 & n.a. & 1/16, 1/4, 1, 4, 16 & 12800 & $64\times100\times64$ & isothermal & n.a. & n.a. & n.a.\\ \hline 
IsoBphi4 & 10000 & n.a. & 1/16, 1/4, 1, 4, 16 & 12800 & $64\times100\times64$ & isothermal & n.a. & n.a. & n.a.\\ \hline 
IsoBphi5 & 10000 & n.a. & 1/16, 1/4, 1, 4, 16 & 12800 & $128\times200\times128$ & isothermal & n.a. & n.a. & n.a.\\ \hline 
IsoBphi6 & 10000 & n.a. & 1/16, 1/4, 1, 4, 16 & 25600 & $128\times200\times128$ & isothermal & n.a. & n.a. & n.a.\\ \hline 
IsoBphi7 & 10000 & n.a. & 1/16, 1/4, 1, 4, 16 & 51200 & $256\times400\times256$ & isothermal & n.a. & n.a. & n.a.\\ \hline 
IsoBz1 & n.a. & $100$ & 1/16, 1/4, 1, 4, 16 & 12800 & $64\times100\times64$ & isothermal & n.a. & n.a. & n.a.\\ \hline 
IsoBz2 & n.a. & 1000 & 1/16, 1/4, 1, 4, 16 & 12800 & $64\times100\times64$ & isothermal & n.a. & n.a. & n.a.\\ \hline 
IsoBz3 & n.a. & 10000 & 1/16, 1/4, 1, 4, 16 & 12800 & $64\times100\times64$ & isothermal & n.a. & n.a. & n.a. \\ \hline 
IsononetB & n.a. & n.a. & 1/16, 1/4, 1, 4, 16 & 12800 & $64\times100\times64$ & isothermal & n.a. & n.a. & n.a. \\ \hline 
&&&&&&&&& \vspace{-0.3cm}\\
IdealBphi1 & 10000 & n.a. & variable (1/16-16) & 25600 & $128\times200\times128$ & ideal & 0.6$\dot{7}$, 0.7$\dot{3}$, 0.8, 0.9$\dot{3}$, 1.0,  1.0$\dot{6}$ & 0.4 & 16\\ \hline 
&&&&&&&&& \vspace{-0.3cm}\\
IdealBphi2 & 10000 & n.a. & variable (1/16-16) & 25600 & $128\times200\times128$ & ideal & 0.6$\dot{7}$, 0.7$\dot{3}$, 0.8, 0.9$\dot{3}$, 1.0,  1.0$\dot{6}$ & 0.4 & 1/16 \\ \hline 
\end{tabular}
\caption{Table showing the simulation parameters used in this work, where $\beta_{x}=(c_{s}/v_{A\,x})^{2}$ is the plasma beta of the initial net magnetic field in the $x$-direction and $\kappa$ is the electron scattering opacity used in the cooling function in cgs units.}
\label{Table1}
\end{table*}
\subsection{Initial conditions}
Explicit parameters of our numerical simulations are provided in Table \ref{Table1}.  All simulations start with a zero net field in the $z$-direction, $B_{z}=B_{0}\sin(2\pi x/H)$ and plasma beta, $\beta_{0}=100$, where $\beta_{0}=2 P/B_{0}^{2}$ is the ratio of thermal to magnetic pressure.  We shall subsequently quantify the strength of any net magnetic field as $\beta_{x}=2 P/B^{2}_{x}$, the ratio of thermal pressure to the magnetic pressure of the net field in the $x$-direction. The dimensionless orbital frequency and thermal sound speed are chosen to be $\Omega=c_{s}=0.001$, so $H=1$, and box lengths are $L_{x}=H$, $L_{y}=4H$ and $L_{z}=H$. Simulations were initialised with random pressure perturbations of maximum amplitude $1\%$, as in \cite{2009ApJ...707..833S}. 
\begin{figure}
	\centering
            \includegraphics[height=5.5cm, angle=0,clip=true, trim=0cm 0cm 0cm 0cm]{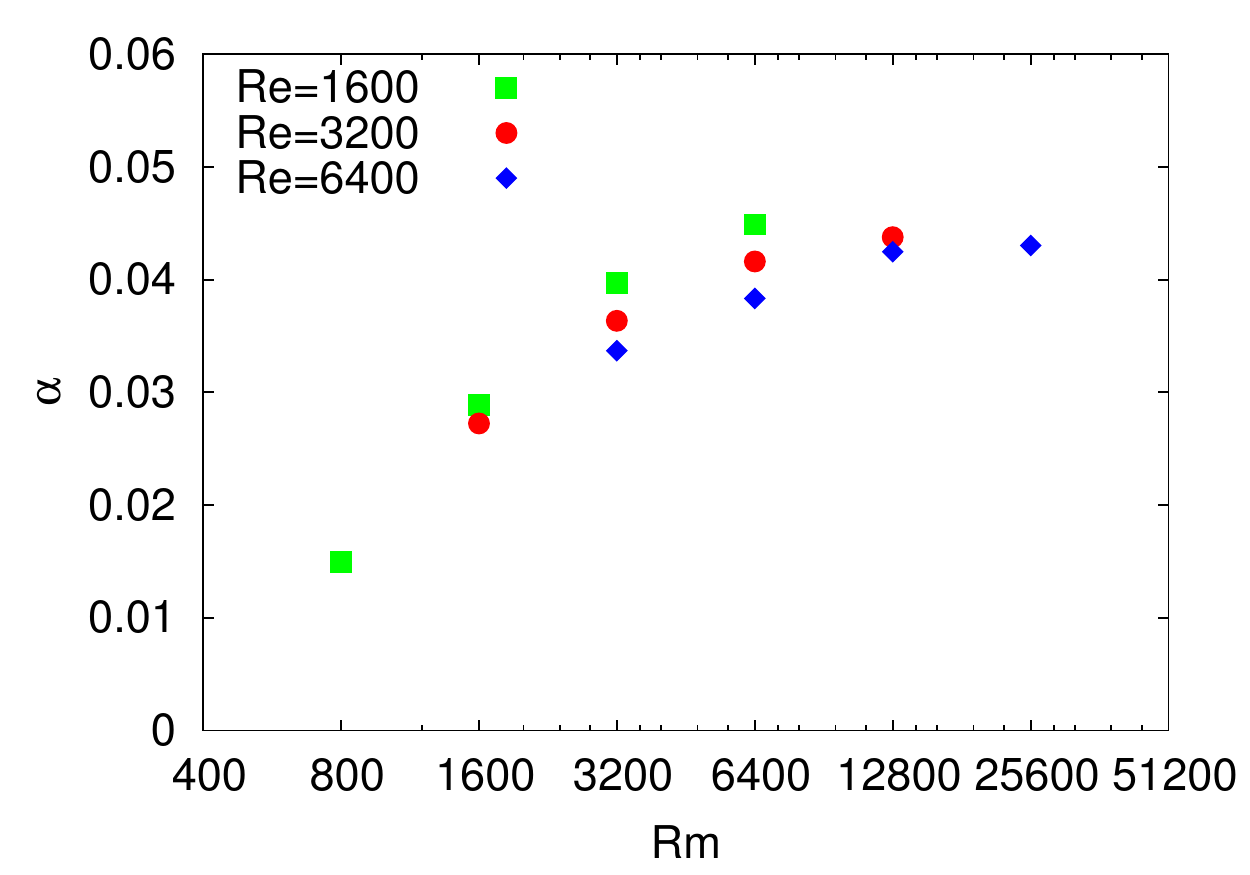} 				
	\caption{Average turbulent disc stress $\alpha$ for different values of Re and Rm with a net B-field in the azimuthal direction $\beta_{\phi}=100$. Large values of the resistivity have a more pronounced effect than either the viscosity or $\Pm$ ($\rm{Rm}/\rm{Re}$), which suggests that a resistivity $\rm{Rm}<1600$ affects the macro-scale linear growth rate of the MRI.
Values of Rm in excess of 12800 appear to exhibit convergence in these low resolution runs (IsoBphi1, Table \ref{Table1}), essentially becoming insignificant compared to the numerical grid resistivity. }
\label{fig1}
\end{figure}
\begin{figure}
	\centering
            \includegraphics[height=5.5cm, angle=0,clip=true, trim=0cm 0cm 0cm 0cm]{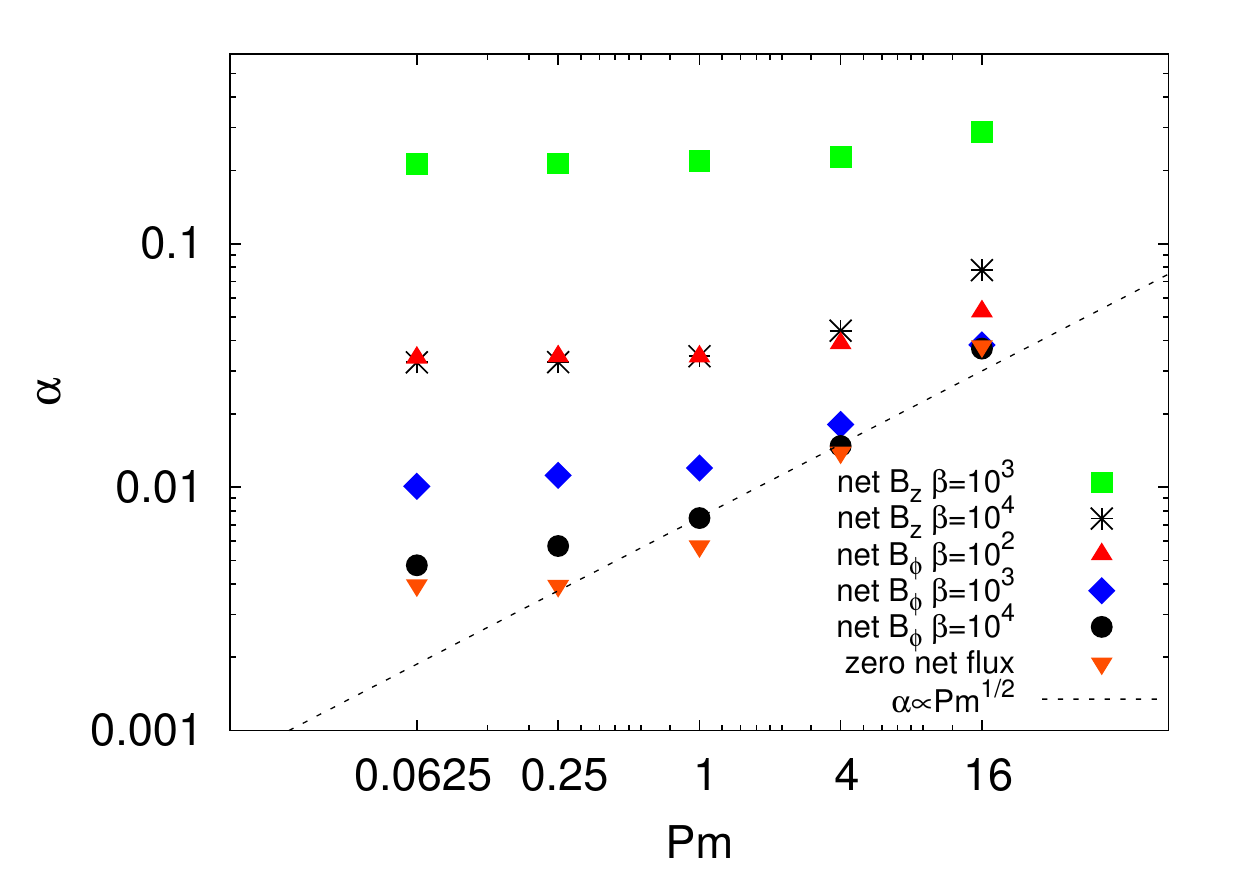} 				
	\caption{Average turbulent disc stress $\alpha$ for different values of the magnetic Prandtl number and net magnetic fields at fixed $\rm{Rm}=12800$. Larger net magnetic field strengths lead to larger values of $\alpha$ in the low $\Pm$ regime. The dashed line shows the instability threshold (\ref{geninst}), $\alpha$-$\Pm$ dependencies with larger gradients than the dashed line are unstable. We expect the instability to occur for low net $B_{\phi}$ fields with $\beta_{\phi}<1000$. Simulation parameters are given in Table \ref{Table1}.}
\label{fig2}
\end{figure}
\begin{figure}
	\centering
            \includegraphics[height=5.5cm, angle=0,clip=true, trim=0cm 0cm 0cm 0cm]{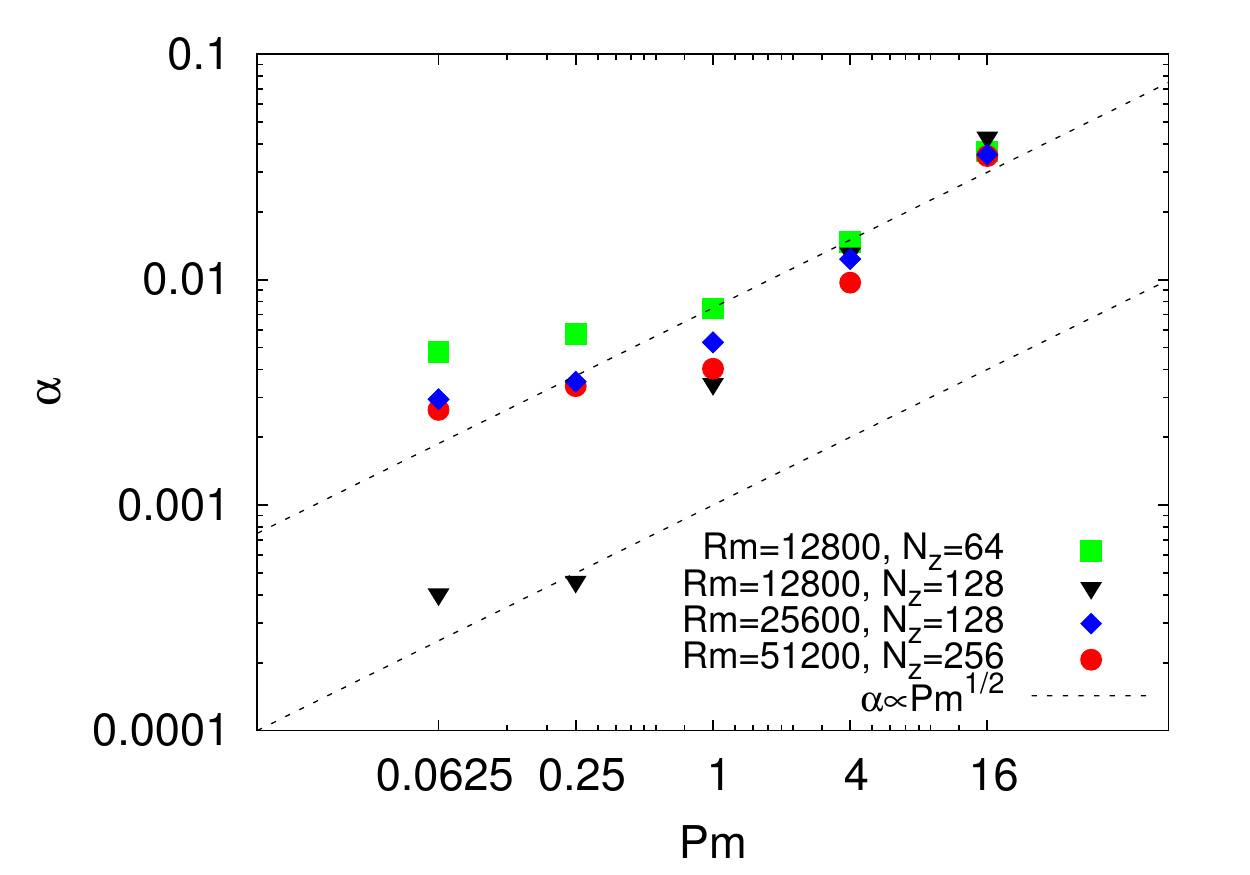} 				
	\caption{The effect of changing numerical resolution and the initial fixed value of $\rm{Rm}$ on the average turbulent disc stress $\alpha$ for a fixed range of values of the magnetic Prandtl number. All these simulations have a net $B_{\phi}$ field with $\beta_{\phi}=10^{4}$ and $N_{z}$ is the number of grid zones per disc scale height, H, in the $z$-direction.  The $\alpha$-$\Pm$ dependence is strong enough in all of these simulations at $\Pm>1$ to lead to a thermal instability (\ref{geninst}), as indicated by possessing a gradient greater than the dotted lines. The results strongly support the hypothesis that there is a real physical $\alpha$-$\Pm$ dependence since this dependence is maintained when simultaneously increasing the simulation resolution and decreasing the resistivity and viscosity (see text for discussion). }
\label{fig3}
\end{figure}

We first investigate disc turbulence for a variety of values of Rm and Re. Our findings are in basic agreement with previous investigations, which found that larger values of viscosity increase the value of the $\alpha$, whilst large resistivities decrease $\alpha$ (Fig.  \ref{fig1}). There is clearly a complex interaction between the viscosity, resistivity, net B-field and resolution. Large values of the viscosity and resistivity, $\rm{Re}<800$, $\rm{Rm}<800$, result in sufficiently high dissipation on large lengthscales to prevent sustainable turbulence. These values are not representative of what would be expected in realistic astrophysical disc environments, and we avoid this regime.
On the other hand, excessively small values of the dissipation coefficients (i.e. large Reynolds numbers) will produce no measurable effect in the simulations, since the relevant lengthscales on which the viscosity and resistivity become important will be below the grid resolution (and unavoidable numerical dissipation) of the simulations. Moreover, whilst an excessively large viscosity {\it reduces} the linear growth rate of the MRI, a large resistivity {\it eliminates} the linear MRI. We focus therefore on the intermediate asymptotic regime in which the resistivity is held constant at the lowest resolvable value, and allow only the viscosity to vary in order to study the effect of altering $\Pm$.

The results of varying $\Pm$ (with fixed Rm) are shown in Figures \ref{fig2} and \ref{fig3} and Table \ref{Table1}. The results, as expected, show an increase in the strength of the saturated MRI turbulence for values of $\Pm>1$.   At small values of $\Pm$, $\alpha$ tends to a constant value determined by the initial net B-field in the simulation, $\rm{Rm}$, and the resolution of the simulation. The presence of an imposed initial net magnetic field is expected to increase the strength of saturated disc turbulence because it provides a sort of  \lq{}backbone\rq{} magnetic field which cannot be dissipated and is thus a permanent source of magnetic field.   The initial net magnetic field is crucial in determining the minimum value of $\alpha$ at small $\Pm$. Net $B_{\phi}$ fields have a strong impact on the $\alpha$ value at $\Pm\lesssim1$, with larger $B_{\phi}$ fields increasing the value of $\alpha$. Net $B_{z}$ fields increase $\alpha$ at all values of $\Pm$ (the strongly enhancing effect of a net $B_{z}$ field has been known for some time \citealt{1995ApJ...440..742H}). 

\begin{figure*}
	\centering
		\subfloat[$\Pm=1/4$]{ \includegraphics[width=8cm, clip=true, trim=0.6cm 3cm 0.5cm 5cm]{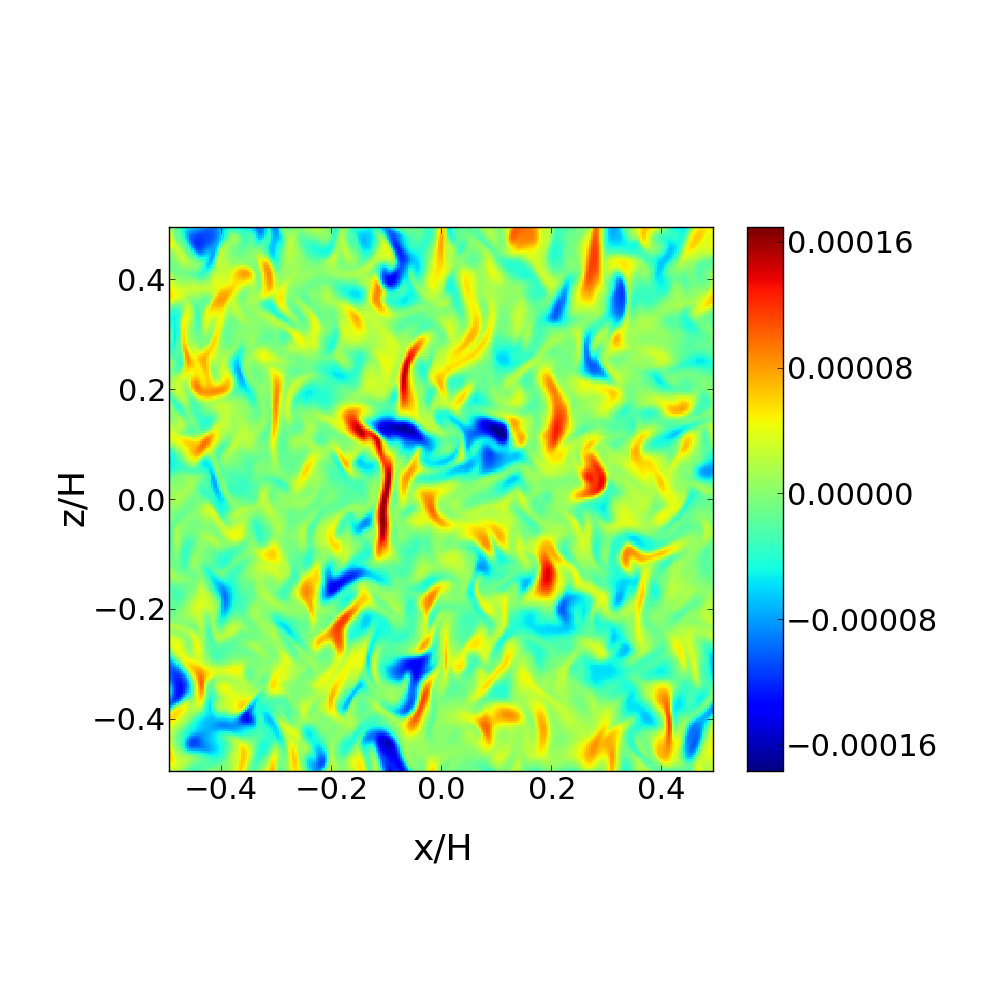} }\qquad
		\subfloat[$\Pm=16$]{ \includegraphics[width=8cm, clip=true, trim=0.6cm 3cm 0.5cm 5cm]{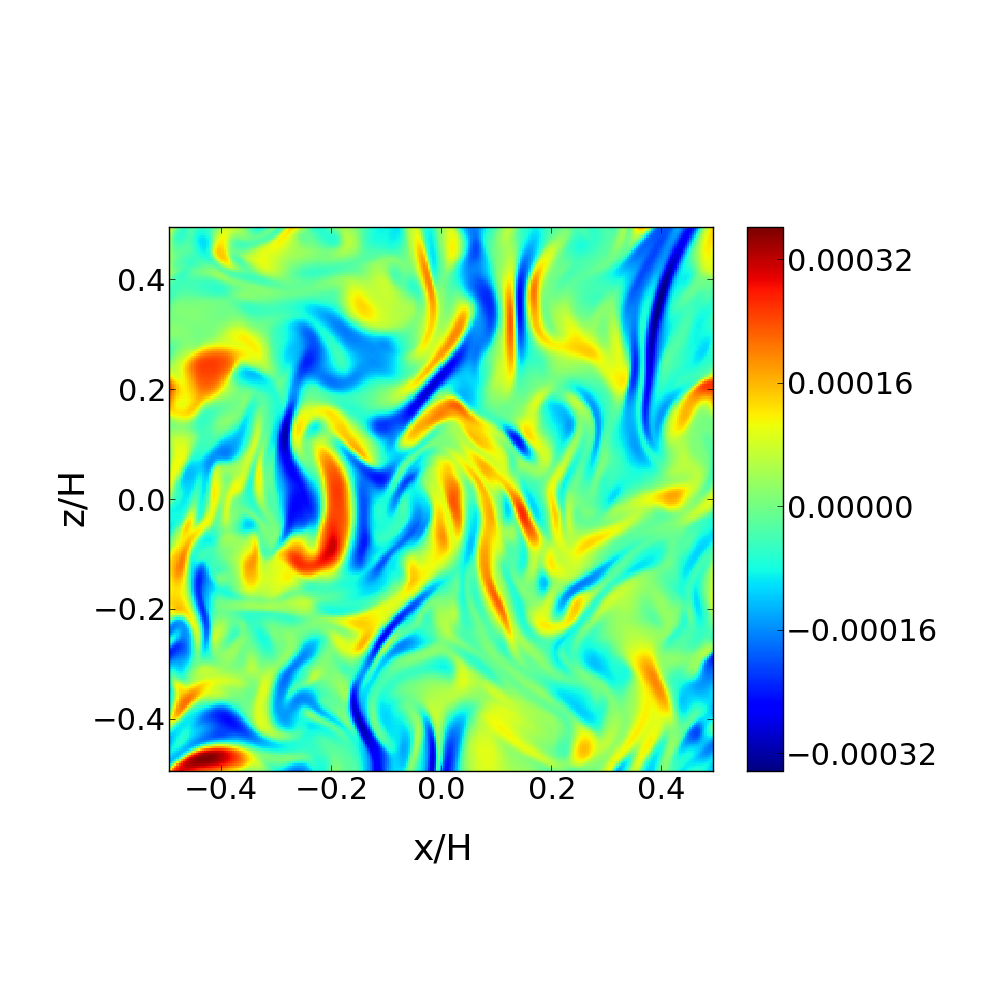} }	
	\caption{Examples of the magnetic field structure in the case of (a): a low $\Pm$ equilibrium solution, (b): a high $\Pm$ equilibrium solution. The radial magnetic field component is shown in an $x$-$z$ slice at the centre of the box for simulations (IsoBphi7) with $\Pm=1/4$ and $\Pm=16$ respectively. The magnetic field is dominated by larger scale structure in (b) with $\Pm=16$, than in (a) with $\Pm=1/4$. This is due to the larger viscosity at larger $\Pm$, which damps small-scale turbulent velocities. This affects the structure of the magnetic field, even though the resistivity is the same in the two simulations. This is evidence in favour of the $\alpha$-$\Pm$ dependence being caused, in part, by a lower reconnection rate at high $\Pm$. The dimensionless magnetic field strength on the colour bar can be compared to the dimensionless isothermal sound speed $c_{s}=0.001$.}
	\label{fig5}
\end{figure*}

The size of the dissipation coefficients is of necessity unphysically large in simulations (due to a lack of dynamic range). To establish the dependence of $\alpha$ on $\Pm$ as a physical, rather than a numerical, effect, we need to show that variations in $\alpha$ exist which depend only upon the {\it ratio} of dissipation coefficients (i.e. $\Pm$) and not on their absolute values.     This is done by comparing the results from simulations covering the same range of $\Pm$ with the same initial conditions, but as the numerical resolution of the simulation is increased, the absolute values of viscosity and resistivity are simultaneously decreased. Results are shown in Figure \ref{fig3} where it can be seen that using the same initial setup, the simulations at higher numerical resolution (which at a given value of $\Pm$ have substantially decreased values of both viscous and resistive dissipation coefficients) obtain remarkably similar results to the lower resolution simulations.  If the $\Pm$-$\alpha$ effect were due to artificially large dissipation lengthscales and insufficient scale separation, we would expect the effect to decrease with increased resolution and decreased values of the dissipation coefficients.  Figure \ref{fig3} shows this is clearly not the case.   In fact, the $\Pm$-$\alpha$ effect is slightly more pronounced for $\Pm<1$ at higher resolution, even when the absolute dissipation coefficients are smaller.  This is a clear indication that whatever residual numerical effects may be present, there appears to be a distinct and genuine physical $\Pm$ effect at work (Fig. \ref{fig5}). 


The formal instability criterion is $\partial \ln \alpha/\partial \ln \Pm>0.5$ (PB). From Figure \ref{fig3} this is satisfied in the $\Pm>1$ regime for the high resolution simulations with small mean magnetic field strengths.  
\begin{figure}
	\centering
            \includegraphics[width=6cm, angle=-90,clip=true, trim=0cm 0cm 0cm 0cm]{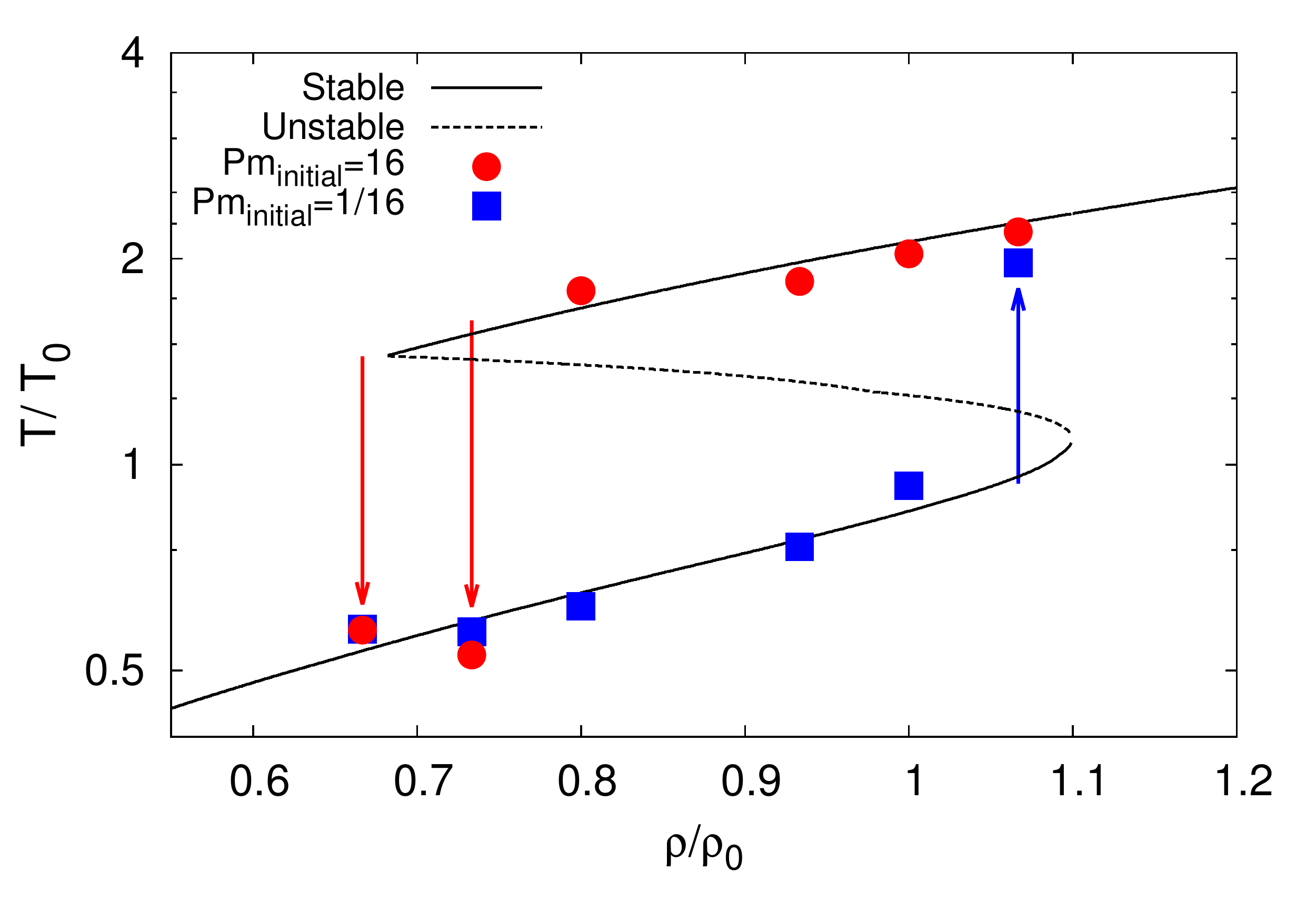} 				
	\caption{Local thermal equilibrium curve showing the two overlapping sets of stable solutions for low and high $\Pm$. The simulations show the characteristic unstable S-curve associated with a thermal-viscous instability. The black lines show the expected thermal equilibrium curve obtained by solving (\ref{Teq}), with solid indicating stable and dashed indicating unstable solutions. This demonstrates the existence of the $\Pm$ instability in a self-consistent set of MHD simulations.}
\label{fig4}
\end{figure}
\begin{figure}
	\centering
            \includegraphics[height=7cm, angle=-90,clip=true, trim=0cm 0cm 0cm 0cm]{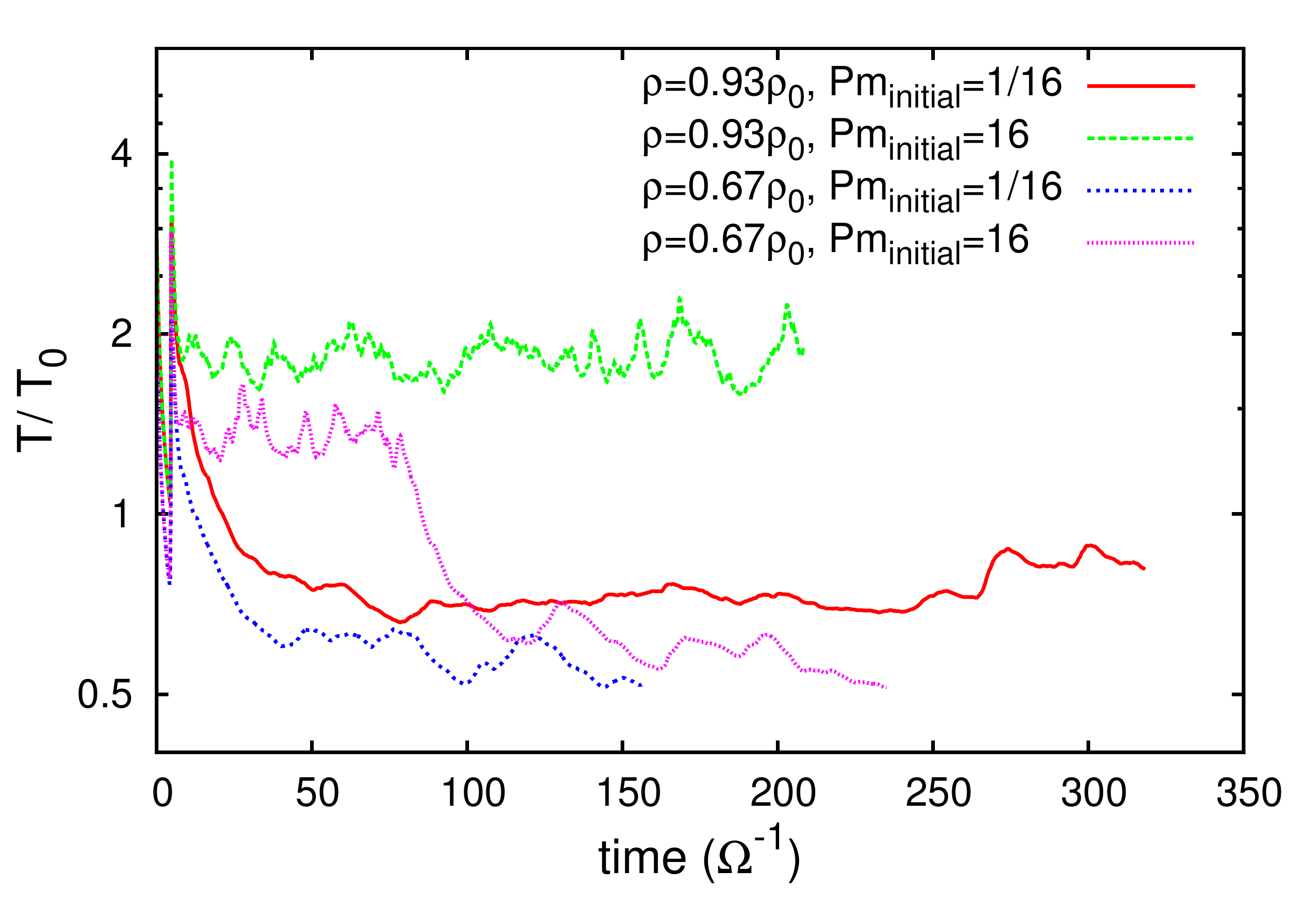} 				
	\caption{The spatially averaged temperature as a function of time for two sets of simulations shown in Fig. \ref{fig4}. The red and green curves show the two stable thermal equilibria which exist at high and low $\Pm$ at a density of $\rho=0.9\dot{3}\rho_{0}$. At a density of $0.\dot{6}\rho_{0}$ only the low $\Pm$ thermal equilibrium solution exists and so both the blue and pink curve converge on the low $\Pm$ solution (after the initial high fixed $\Pm$ value is freed at $\approx 80\Omega^{-1}$ for the simulation shown in pink).}
\label{fig6}
\end{figure}

\section{Thermal equilibrium curve}

Let us now investigate whether the dependence of $\alpha$ on $\Pm$ leads to the thermal-viscous instability outlined in PB. To test this,  the local thermal-equilibrium curve of the disc is simulated. It is necessary to include both turbulent heating and radiative cooling (the gas is no longer isothermal), as well as a variable magnetic Prandtl number which has the correct dependence on temperature and density calculated in (\ref{Pm}).   To minimise the direct effect of a large resistivity on the linear MRI growth rate, we adopt a constant low value for the resistivity ($\rm{Rm}\simeq25600$) and include the temperature and density dependence of $\Pm$ in the viscosity alone. The explicit resistivity and viscosity are given by
\be
\eta=\frac{c_{s0}H_{0}}{25600}, \qquad \nu=\frac{c_{s0}H_{0}}{25600}\left(\frac{T}{T_{0}}\right)^{11/2} .\left(\frac{\rho}{\rho_{0}}\right)^{-2},
\label{Re}
\ee
where in the formula above $c_{s0}$ and $H_{0}$ are the initial values of the thermal sound speed and disc scale height. $T_{0}$ and $\rho_{0}$ are normalisation parameters which are chosen for convenience to be of order unity in Figure \ref{fig4}. The range of allowed $\Pm$ values is limited to $1/16>\Pm>16$ (or $409600>\rm{Re}>1600$). This range is chosen because when $\Pm<1/16$ the viscosity is completely sub-dominant to the numerical grid viscosity and when $\Pm>16$ the viscosity becomes unrealistically large, with the viscous lengthscale becoming comparable to the total size of the simulation. Instability, if present, will manifest itself as two sets of overlapping quasi-stable solutions (stable on the thermal timescale but evolving on the mass accretion timescale, hence the term thermal-viscous instability). To find such solutions, we carry out two simulations with nearly identical initial conditions, the sole difference being that one simulation starts with a fixed high $\Pm=16$ and one with a fixed low $\Pm=1/16$.  The MRI is allowed to grow, become fully turbulent, and attain a dynamical-thermal equilibrium in the two cases.  From Fig. \ref{fig2} the high $\Pm$ simulation is expected to have stronger turbulent heating due to the larger $\alpha$ value, and so to reach a higher equilibrium temperature than the low $\Pm$ simulation. After equilibrium has been achieved, $\Pm$ is freed and its value is calculated via equation \ref{Re} for each grid cell. Thus $\Pm$ becomes a function of the local fluid temperature and density. The two simulations are then evolved until they relax to a new quasi-stable thermal equilibrium, if this in fact differs from the initial equilibrium. If the instability is present in the simulation, this should manifest itself as two stable thermal equlibria at the same surface density $\Sigma$, producing the characteristic unstable S-curve, as in the case of the hydrogen ionisation instability in dwarf novae (e.g. \citealt{2001NewAR..45..449L}, \citealt{2002apa..book.....F} and PB). 

Figure \ref{fig4} shows the local thermal equilibrium curve of the instability calculated using a set of simulations with a weak net magnetic field in the azimuthal direction ($\beta_{\phi}=10^{4}$). It shows that two stable thermal equilibria exist at the same density in the unstable range of $\Pm$ ($\Pm \approx 1-16$). {\it This is the first self-consistent MHD simulation demonstrating the $\Pm$ instability from first principles.} It shows that the dependence of $\alpha$ on $\Pm$ in a full MHD simulation is sufficiently sensitive to induce a thermal-viscous instability in the disc plasma. The black curve in the figure is the expected theoretical thermal equilibrium curve calculated by balancing turbulent heating and radiative cooling. This is in close agreement with the results of the simulation. The black curve is calculated from the thermal equilibrium equation 
\be
\frac{\sigma T_{e}^{4}}{H}=\frac{3}{2}\Omega\alpha(\Pm) P_{\m{tot}},
\label{Teq}
\ee
where the LHS corresponds to radiative cooling and the RHS to turbulent heating (see e.g. \citealt{1998RvMP...70....1B}). The value of alpha is determined as a function of $\Pm$ for the black curve by using the $\alpha$-$\Pm$ dependence found in simulation IsoBphi6, which is the equivalent isothermal simulation. Due to the finite resolution of the simulation, turbulent fluctuations in the spatially averaged values of $\alpha$ and $T$ are much larger than would be expected in an actual disc (in part, because of Poisson noise from the finite number of grid points in the simulation) and so this precludes accurately simulating the solutions close to the edges or corners of the solid S-curve.  Close to the edges of the upper and lower branches of solutions small temporal fluctuations in temperature are sufficient to move the average temperature of the plasma over the dashed line of solutions which are unstable to perturbations. The dashed line represents the watershed between states which will experience runaway heating (above the dashed line) or cooling (below the dashed line) until they reach an equilibrium solution on the solid curve. Examples of the evolution of thermally stable and unstable simulations are shown in Figure \ref{fig6}.  The precise properties of the thermal-equilibrium curve are clearly dependent on the initial net magnetic field, which is likely to vary between different astrophysical systems. 

\section{Conclusion}

In this paper we address two important questions - (i) to what extent does the turbulent disc stress depend on the magnetic Prandtl number, and is this effect numerical or physical? (ii) is the $\alpha$-$\Pm$ dependence sufficiently sensitive to trigger a thermal-viscous instability in the disc? 

In the first section of the paper we address (i) by conducting a suite of isothermal 3D MHD local shearing-box simulations to investigate the dependence of the turbulent disc stress $\alpha$ on the viscous and resistive dissipation coefficients. In agreement with previous studies we find that $\alpha$ depends on the magnetic Prandtl number ($\Pm$), the ratio of viscosity to resistivity. In the regime $1<\Pm<16$, $\alpha$ increases with $\Pm$ and the dependence is sufficiently sensitive to induce an important new thermal-viscous instability, the magnetic Prandtl number disc instability (see PB). We investigate whether the $\alpha$-$\Pm$ dependence is physical or numerical in origin by conducting a suite of simulations covering the same range in $\Pm$, but at different numerical resolutions and with different absolute values of viscosity and resistivity. If the $\alpha$-$\Pm$ dependence were numerical and caused by artificially large dissipation coefficients or a lack of scale separation in the turbulent cascade, the dependence should decrease as the simulation resolution is increased and the dissipation coefficients are decreased. In fact, the $\alpha$-$\Pm$ dependence persists as the simulation resolution is increased and the absolute values of the dissipation coefficients are decreased.  This shows that the $\Pm$-$\alpha$ effect is not a numerical artefact caused by artificially large dissipation coefficients or a lack of scale separation. It is firm evidence that the $\alpha$-$\Pm$ dependence is physical in origin. 

To investigate whether the $\alpha$-$\Pm$ dependence is sufficiently sensitive to induce the magnetic Prandtl number disc instability we conducted a further set of MHD simulations which include realistic turbulent heating and radiative cooling to map out the local thermal-equilibrium curve of the disc. In these simulations $\Pm$ is no longer held fixed; instead it is given a realistic physical dependence on the plasma density and temperature. Significantly, we find that the $\alpha$-$\Pm$ dependence is sufficiently strong to trigger the $\Pm$ instability using physical conditions appropriate for an accretion disc. The thermal-equilibrium curve is found to have a characteristic S-shape forming an unstable limit cycle. In the unstable region corresponding to the regime $1<\Pm<16$, the instability manifests itself as two sets of overlapping stable solutions with different equilibrium temperatures but the same surface density. The two sets of stable solutions correspond to a hotter, high $\alpha$ branch of solutions with large $\Pm$ and a cooler, low $\alpha$ branch of solutions corresponding to $\Pm<1$ as predicted in PB. This is the first self-consistent MHD simulation demonstrating the existence of the $\Pm$ disc instability from first principles. It was shown in PB that the local $\Pm$ instability leads to the production of global heating and cooling fronts propagating through the disc, resulting in changes to the disc state on timescales substantially shorter than those of the DIM. This makes the magnetic Prandtl number disc instability a good potential candidate to explain the hard/soft changes in disc state observed on week long timescales. A detailed comparison of the expected timescale and spectral properties of the global instability will be the subject of future work.



\section{Acknowledgements}

WJP is supported by a Junior Research Fellowship from University College, University of Oxford. SAB acknowledges support from the Royal Society in the form of a Wolfson Research Merit Award. We acknowledge support from STFC in the form of a Consolidated Grant to Oxford Astrophysics. The simulations presented here were run on the DiRAC Complexity Cluster at Leicester and on Berg, the DiRAC facility jointly funded by STFC, the Large Facilities Capital Fund of BIS and the University of Oxford. WJP is particularly grateful to Julien Faure for important numerical advice and to Sebastien Fromang for hosting a visit to the CEA Saclay. We would like to thank Sebastien Fromang, Julien Faure, Charles Gammie, Rob Fender, John Miller, Jean-Pierre Lasota and Alexander Schekochihin for helpful conversations and suggestions.

\bibliographystyle{mn2e}
\bibliography{JRF}
\bibdata{JRF}

\label{lastpage}

\end{document}